%% file: jetc-article-final.tex
\documentclass[acmsmall]{acmart}

\AtBeginDocument{%
  \providecommand\BibTeX{{%
    \normalfont B\kern-0.5em{\scshape i\kern-0.25em b}\kern-0.8em\TeX}}}

\setcopyright{acmcopyright}
\acmJournal{JETC}
\acmYear{2022} \acmVolume{1} \acmNumber{1} \acmArticle{1} \acmMonth{1} \acmPrice{15.00}\acmDOI{10.1145/3517812}

\usepackage{graphicx}
\usepackage{hyperref}
\usepackage{amsmath}
\usepackage{color} 
\usepackage{array}
\usepackage{xspace}
\hypersetup{
    colorlinks=true,
    linkcolor=blue,
    filecolor=magenta,      
    urlcolor=cyan,
}
\usepackage{multicol}
\usepackage{caption}
\usepackage{subcaption}
\usepackage{float}

\usepackage[normalem]{ulem}
\usepackage{xcolor,colortbl}
\usepackage{tablefootnote}

\usepackage[T1]{fontenc}

\newcommand{\tbm}[1]{{\ttfamily #1}\xspace}
\newcommand{\ie}{\tbm{IE}} 
\newcommand{\nvr}{\tbm{NVR}} 

\definecolor{codegreen}{rgb}{0,0.6,0}
\definecolor{codegray}{rgb}{0.5,0.5,0.5}
\definecolor{codepurple}{rgb}{0.58,0,0.82}
\definecolor{backcolour}{rgb}{0.95,0.95,0.92}

\newcommand{\sysname}{NORM\xspace}
\newcommand{\signal}[1]{{\ttfamily #1}}

\begin{document}

\title{\sysname: An FPGA-based Non-volatile Memory Emulation Framework for Intermittent Computing}

\author{Simone Ruffini}
\affiliation{%
  \institution{University of Trento}
  \streetaddress{Department of Information Engineering and Computer Science, Via Sommarive, 9, Povo, 38123 TN}
  \city{Trento}
  \country{Italy}}
\email{simone.ruffini@studenti.unitn.it}

\author{Luca Caronti}
\affiliation{%
  \institution{University of Trento}
  \streetaddress{Department of Information Engineering and Computer Science, Via Sommarive, 9, Povo, 38123 TN}
  \city{Trento}
  \country{Italy}}
\email{luca.caronti@studenti.unitn.it }

\author{Kas{\i}m Sinan Y{\i}ld{\i}r{\i}m}
\affiliation{%
  \institution{University of Trento}
  \streetaddress{Department of Information Engineering and Computer Science, Via Sommarive, 9, Povo, 38123 TN}
  \city{Trento}
  \country{Italy}}
\email{kasimsinan.yildirim@unitn.it}

\author{Davide Brunelli}
\affiliation{%
  \institution{University of Trento}
  \streetaddress{Department of Industrial Engineering, Via Sommarive, 9, Povo, 38123 TN}
  \city{Trento}
  \country{Italy}}
\email{davide.brunelli@unitn.it}

\include{0-abstract}

\maketitle 

\input{1-introduction}
\input{2-background}
\input{3-system}

\input{4-implementation}

\input{5-evaluation}
\input{6-conclusions}

\bibliographystyle{ACM-Reference-Format}
\bibliography{jetc-article-final}

\end{document}

%% file: 0-abstract.tex
\begin{abstract}
Today's intermittent computing systems operate by relying only on harvested energy accumulated in their tiny energy reservoirs, typically capacitors. An intermittent device dies due to a power failure when there is no energy in its capacitor and boots again when the harvested energy is sufficient to power its hardware components. Power failures prevent the forward progress of computation due to the frequent loss of computational state. To remedy this problem, intermittent computing systems comprise built-in fast non-volatile memories with high write endurance to store information that persists despite frequent power failures. However, the lack of design tools makes fast-prototyping these systems difficult. Even though FPGAs are common platforms for fast prototyping and behavioral verification of continuously-powered architectures, they do not target prototyping intermittent computing systems. This article introduces a new FPGA-based framework, named \sysname (\textbf{N}on-volatile mem\textbf{OR}y e\textbf{M}ulator), to emulate and verify the behavior of any intermittent computing system that exploits fast non-volatile memories. Our evaluation showed that NORM can be used to emulate and validate FeRAM-based transiently-powered hardware architectures successfully.
\end{abstract}

\ccsdesc[500]{Hardware~Energy generation and storage}
\ccsdesc[500]{Computer systems organization~Embedded hardware}
\ccsdesc[500]{Hardware~Reconfigurable logic applications}
\ccsdesc[500]{Computer systems organization~Special purpose systems}

\keywords{Intermittent computing, Non-volatile Processors, FPGAs.}

%% file: 1-introduction.tex
\section{Introduction}\label{sec:introduction}

The recent advancements in microelectronics led to the emergence of batteryless sensors that operate relying only on ambient energy~\cite{sample2008design}. This sensing technology opens up new application spaces where small devices should have eternal lifetimes,  autonomous operation, and massive deployments in inaccessible locations~\cite{hester2017future}. Batteryless sensors comprise energy harvesting circuits that use several sources such as solar, thermal, and radio waves to accumulate the environmental energy into a small energy buffer, typically a tiny capacitor. When the stored harvested energy is above an operating threshold, the microcontroller reboots to compute, sense, and communicate. When the energy drains out of the capacitor, the microcontroller and peripherals turn off due to a power failure. Today's batteryless sensors are composed of ultra-low-power microcontrollers whose main architectural components are \emph{volatile}. When the batteryless sensor turns off due to a power failure, the volatile processor state (e.g., the contents of the stack, program counter, registers) is lost. This operation leads to the loss of all computational states and intermediate results~\cite{ransford2012mementos}.

The software on batteryless platforms runs \emph{intermittently} due to frequent charge-discharge cycles. As a consequence of the intermittent execution, the computation might not \emph{progress forward} and \emph{memory consistency} might be violated~\cite{ransford2014nonvolatile}. To store information that will persist despite power failures, microcontrollers in batteryless sensors comprise embedded non-volatile secondary memory components, e.g., FeRAM~\cite{FeRAM,FeRAMDatasheet} that exhibits low-power characteristics, faster write performance and greater maximum read/write endurance compared to Flash memories, even if they pay lower memory density than other recent NVM technologies. Using software-based techniques (e.g., ~\cite{balsamo2016hibernus++,yildirim2018ink,kortbeek2020time}), programmers backup the volatile state of the microcontroller into non-volatile memory to recover computation from where it left upon a power failure. As an alternative to software-based solutions, leveraging non-volatile logic and building non-volatile processors (NVPs) is another approach to ensure forward progress of computation and keep memory consistent during intermittent execution.  NVPs integrate built-in non-volatile memory in their architecture. They automatically back up the computation state into their internal non-volatile registers upon a power failure and restore the state upon recovery~\cite{ma2016nonvolatile}. All these operations are transparent to the programmer.   

The architectural design space of intermittent computing systems that exploit non-volatile logic is broad and includes several design options with different pros and cons. As an example, a crucial design decision is to identify which state elements will be non-volatile. Systems designers can keep all registers non-volatile,  which is slower and more energy-consuming. Alternatively, the designers can keep all registers as volatile, but they can maintain additional non-volatile registers to back up the volatile state (i.e., volatile registers) at specific points in time. Another crucial issue is to decide the backup frequency of the volatile state components. For instance, a computing system can backup its state at every clock cycle, or it can backup on-demand ~\cite{ma2015architecture,ma2016nonvolatile}, to decrease the backup frequency and save energy. However, the lack of design tools makes fast-prototyping and functional verification of computing systems with non-volatile logic difficult. 
FPGAs (field-programmable gate arrays) are useful for fast prototyping and verification of digital logic. As of now, FPGA  fabrics include logic elements that are implemented using either volatile memory or non-volatile memory~\cite{tang2018post}, but not both. Therefore, existing HDLs (such as VHDL or Verilog) do not provide specific keywords to make a differentiation between a volatile state element and a non-volatile state element. This situation prevents hardware designers from using  FPGAs to fast-prototype their logic designs targeting intermittent computing, which include both volatile and non-volatile logic. To the best of our knowledge, the state-of-the-art does not propose a solution to emulate transiently-powered intermittently operating hardware architectures using off-the-shelf FPGAs.  

In this article, we introduce a new FPGA-based framework,  named \sysname (\textbf{N}on-volatile mem\textbf{OR}y e\textbf{M}ulator), that can be used to emulate any intermittent computing system with fast non-volatile memory. \sysname can be used to debug and perform functional verification of non-volatile computing logic. Moreover, \sysname can be integrated into a working intermittent computing system in place of a yet-to-be-built non-volatile computing logic so the whole system can be tested. \sysname comprises auxiliary blocks that:
\begin{enumerate}
    \item simulates the behavior of irregular power supply typical to energy-harvesting intermittent systems; 
	\item simulates the persistence of the non-volatile micro-architectural elements as well as the long delay of reading/write operations (as compared to those of volatile memory);
	\item approximates the power consumption of the emulated technology. 
\end{enumerate}
Our simulations showed that NORM can be used to emulate and validate FeRAM-based transiently-powered hardware architectures successfully. We release the source code of \sysname (implemented in VHDL) in a public repository~\cite{norm-github} to increase the impact of this work and enable the community to fast prototype and validate transiently-powered non-volatile hardware architectures.

The rest of this article is organized as follows. In Section~\ref{sec:background}, we provide the related work on intermittent computing and NVPs. We present the general description of \sysname in Section~\ref{sec:prop_emu_arc}. Section~\ref{sec:implementation} presents the implementation details of \sysname and Section~\ref{sec:evaluation} presents our evaluation based on simulations. Finally, Section~\ref{sec:conclusion} concludes our article and proposes future work.

%% file: 2-background.tex
\section{Background and Related Work}
\label{sec:background}

\begin{figure}
	\includegraphics[width=\columnwidth]{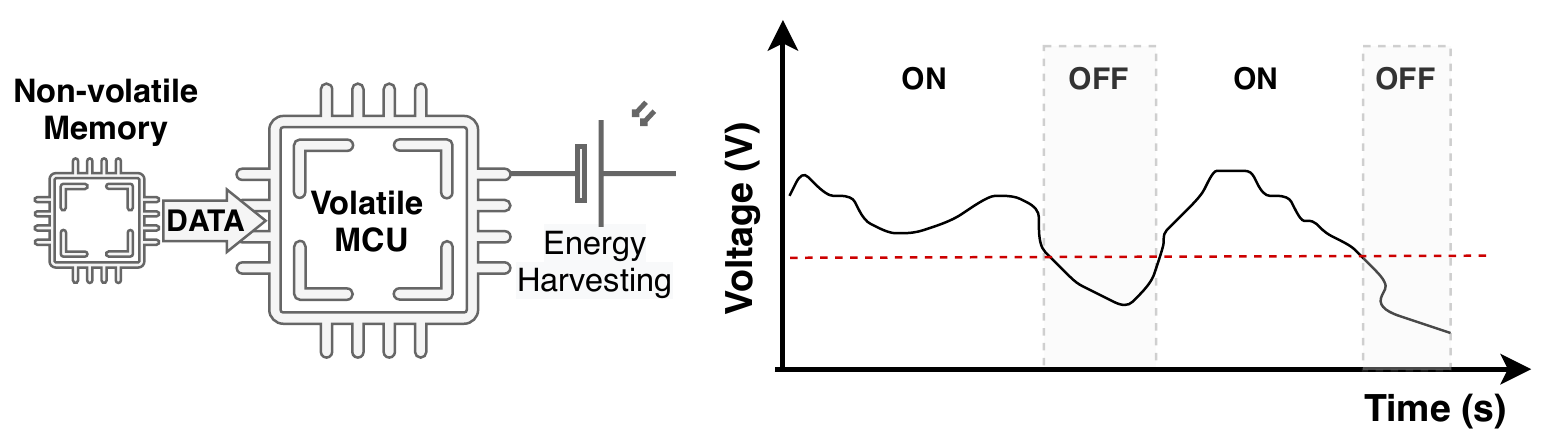}
	\caption{This figure presents the operation of an energy harvesting battery-less computing device. The device harvests energy and stores it in a capacitor. The device dies when the voltage level of the capacitor is below a threshold. The device boots again when the voltage level is above this threshold. This operation leads to intermittent computation. The volatile micro-controller is equipped with external non-volatile memory to store intermediate results and computation state to recover from power failures.}
	\label{fig:intermittent}
\end{figure}

A new class of embedded devices that can sense, compute, and communicate without batteries emerged. As an example, RF-powered batteryless sensors~\cite{Torrisi2020TRAP,gollakota2013emergence} solely rely on the harvested energy of ambient radio frequency waves in the air (see Figure~\ref{fig:intermittent}).  
These batteryless devices, which can feature even more complex sensors such as cameras~\cite{Nardello2019Camaroptera}, comprise ultra-low-power microcontrollers (e.g., MSP430FR5969~\cite{EXP430FR5969_website}) whose main architectural components, e.g., registers and main memory, are volatile. These volatile processors include also a non-volatile secondary memory, e.g., Ferroelectric RAM (FRAM)~\cite{FRAM}, to store information that will persist upon power failures. Despite several ultra-low-power operation modes of these microcontrollers (e.g., sleep mode requires current on the order of a few $\mu$A), batteryless sensors cannot be available continuously using unreliable and sporadic energy sources~\cite{adegbija2018microprocessor,shafik2018real}. Frequent and unpredictable power failures reset the volatile state of the device, prevent the forward progress of computation and hinder its memory consistency. Therefore, programs and libraries designed for continuously-powered computers cannot run on batteryless sensors correctly due to the frequent loss of volatile state, and in turn, failed computation.

\subsection{Intermittent Computing with Volatile Processors}

Volatile microcontrollers employ \emph{software-aided} solutions to
mitigate the effects of unpredictable power failures. Generally speaking, these solutions backup the volatile state of the processor into non-volatile memory to recover computation from where it left upon a power failure. Moreover, they ensure memory consistency so that the backed-up state in the non-volatile memory will not be different than the volatile state, or vice versa. Current literature proposed mainly two software-based approaches. One approach is to store the computation state in non-volatile memory via \emph{checkpoints} paired with C programs~\cite{ransford2012mementos,jayakumar2015quickrecall,balsamo2016hibernus++,van2016intermittent,hicks2017clank,lucia2015simpler,bhatti2017harvos,maeng2018adaptive,kortbeek2020time}. Upon recovery from a power failure, the computation continues from the consistent volatile state stored in the latest successful checkpoint. Another approach is to use a custom \emph{task-based} programming model to develop intermittent applications, which eliminates the high cost of checkpointing~\cite{colin2016chain,hester2017timely,yildirim2018ink,maeng2019alpaca,majid2020dynamic,ruppel2019transactional,maeng2020adaptive}. In this model, programmers implement the applications as a collection of idempotent and atomic tasks by employing an explicit task-based control flow. Individual task sizes should not exceed the capacity of the capacitor to ensure forward progress. 
However, software-aided recovery solutions require transmitting data from built-in volatile components of the processor, e.g., registers, to non-volatile memory. This operation suffers from low speed, e.g., 200 $\mu$s~\cite{shi2018time}, and a large energy penalty that grows with the size of volatile elements~\cite{su2017nonvolatile,shi2018time}. Moreover, these solutions require programmers to structure their software by considering programming models designed for intermittent systems, e.g, task-based programming~\cite{colin2016chain,yildirim2018ink}. 

\subsection{Non-volatile Logic and Processors}
Non-volatile processors (NVPs) bring non-volatile memory into the micro-architecture of the processor. Non-volatile logic enables the backup and recovery operations from a power failure to be transparent to the programmer. Moreover, backup and recovery introduce less overhead than software-aided solutions, e.g., only on the order of a few $\mu$s~\cite{wang20123us,li2017advancing}. Since backup and retention operations are fast as compared to the software-aided solutions, NVPs reduce leakage power by shutting down the system when the device  idle~\cite{adegbija2018microprocessor}. 

Due to the higher power required for non-volatile memory read/write operations, NVPs might also consume more power as compared to volatile processors~\cite{ma2015architecture}. Therefore, there is room for micro-architecture-level optimizations to reduce their energy consumption. To decrease the energy requirements of NVPs, recent works proposed:
\begin{enumerate}
    \item using more efficient memory technologies, e.g., ReRAM~\cite{liu20164} and hybrid CMOS/ferroelectric non-volatile flipflop~\cite{su2017ferroelectric}; 
    \item  embedding non-volatility into the computing logic, e.g., transistor level, using NCFET~\cite{li2017advancing} so that logic gates could also store their states intrinsically in a non-volatile fashion; 
    \item using new backup strategies, e.g., backup at every processor cycle or on-demand backup~\cite{ma2015architecture,ma2016nonvolatile}, to decrease backup frequency to save energy.
\end{enumerate} 
These efforts provide implementation technology-level energy optimizations. However, as of now, we do not have tools to fast prototype non-volatile logic and observe optimization strategies targeting different non-volatile intermittent computing architectures and processors. 

\subsection{Non-volatile Memory Simulation/Emulation Frameworks}

There are studies, e.g.,~\cite{zhu2017building,duan2018hme,lee2014fpga,omori2019performance}, that provide the emulation of different non-volatile main memory technologies. These studies present techniques to assess the system's performance concerning different non-volatile memory technologies. Unfortunately, they do not apply to intermittent computing systems. There are studies, e.g.,~\cite{gu2016nvpsim, poremba2012nvmain}, that proposed simulators for non-volatile memory and logic. As an example, NVPSim~\cite{gu2016nvpsim} can simulate the architectural components forming a non-volatile processor by allowing users to select only different configurations (e.g., cache size and organization) for the main high-level components in a processor. Fused~\cite{Fused20} is designed to assess the performance of intermittent systems by providing simulations only at a high level of abstraction. Authors in~\cite{Wu2018} describe a system-level simulator supporting flexible energy behavior configuration for both the processor and peripherals. 

Contrarily to the mentioned studies, this work enables the assessment of non-volatile features in any digital hardware design that includes a combination of volatile and non-volatile logic. We provide full flexibility for the users to evaluate, validate and fast-prototype any HDL design targeting intermittent computing systems. 

\subsection{Field-programmable Gate Arrays (FPGAs)}

FPGAs are used in many applications due to the increased cost and time associated with the custom ASIC (application-specific integrated circuit) design. FPGAs are useful for fast prototyping custom processor architectures and their behavioral verification. Several popular volatile processor architectures, such as RISC-V, have implementations using popular hardware description languages (HDLs) (e.g., Verilog) that can run on FPGAs. OpenFPGA framework~\cite{tang2020openfpga} opened the door for automating the design, verification, and layout of different FPGA architectures. OpenFPGA enabled end-users to port their designs to any FPGAs that OpenFPGA can support. Some recent studies target reducing the energy consumption of FPGAs. As an example, the authors in~\cite{tang2018post} proposed an RRAM-based FPGA architecture, which is inherently fully non-volatile. They replaced the SRAM-based circuits in FPGA architectures with RRAM-based implementations. RRAM-based FPGAs can be powered off during sleep mode and instantly powered on when needed. This strategy reduces the energy requirements.

Using FPGAs to prototype intermittent computing architectures is an open issue. Hardware designs that operate intermittently are composed of volatile and non-volatile logic elements. Current FPGAs provide either volatile or non-volatile state elements, but not both. We do not have mixed memory volatility in the fabric of FPGA architectures. Hence, there are no specific HDL keywords to differentiate a non-volatile register from a volatile one. As of now, hardware designers cannot represent hardware that operates intermittently by using existing HDLs. They cannot validate their designs through simulations. This work focuses on these deficiencies and fills the existing gap in the literature by proposing a novel framework that facilitates the design and validation of intermittently-operating hardware. 

%% file: 3-system.tex
\begin{figure*}

\centering
	\includegraphics[width=\textwidth]{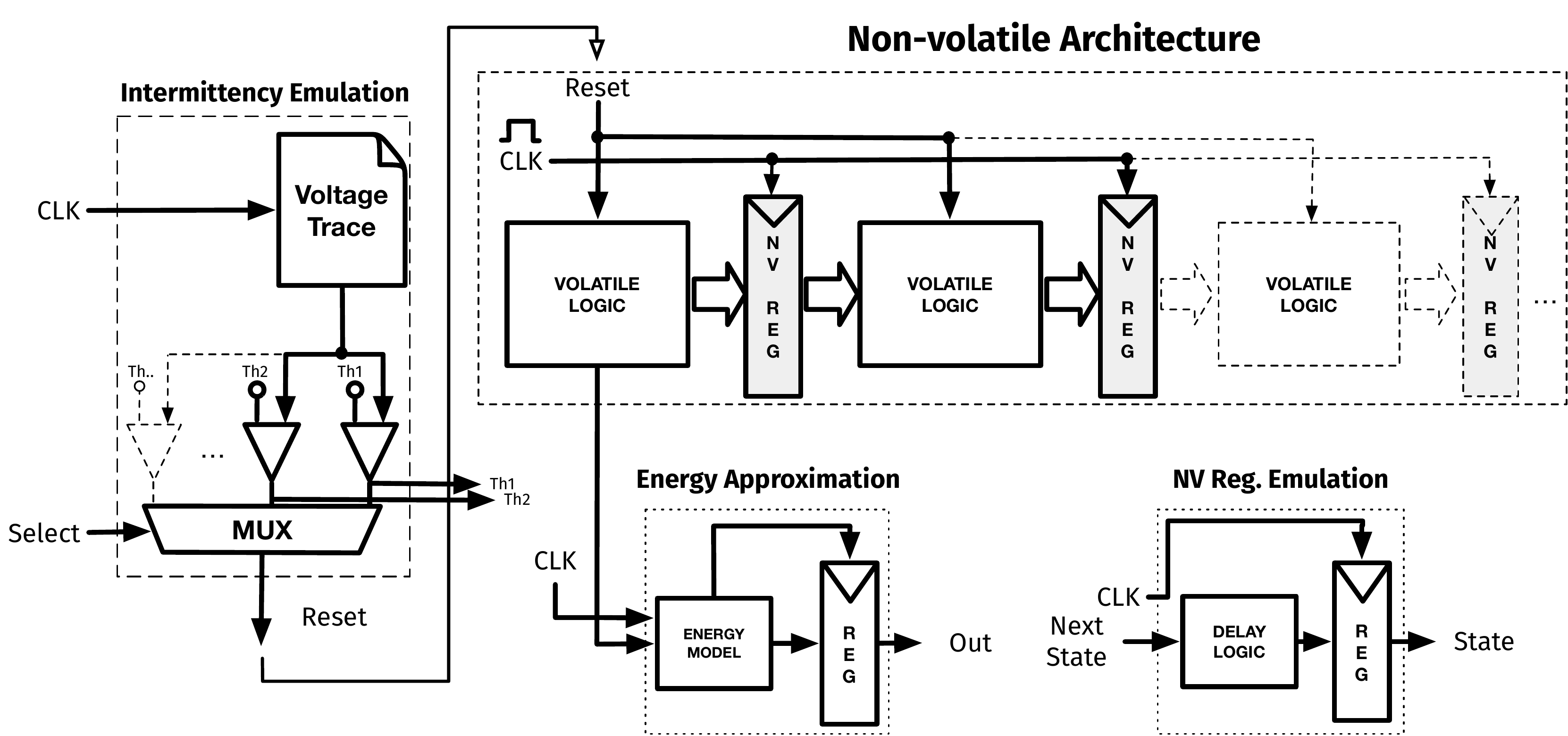}
	\caption{Non-volatile Logic Emulation. Our architecture is composed of three auxiliary blocks. \emph{Intermittency Emulation} emulates intermittent power supply. \emph{Non-volatile Register Emulation} emulates the persistence of the non-volatile registers and the delays of the read/write operations. \emph{Energy Approximation} approximates the energy consumption of the emulated technology.}
	\label{fig:system}
\end{figure*}

\section{\sysname System Overview}
\label{sec:prop_emu_arc}

We propose an emulation architecture, named \textbf{\sysname} (\textbf{N}on-volatile mem\textbf{OR}y e\textbf{M}ulation), that mimics non-volatile memory elements and power failures. \sysname architecture is composed of three main auxiliary blocks: 
\begin{enumerate}
   	\item \emph{Intermittency Emulation} that emulates irregular power supply typical to energy-harvesting systems; 
	\item \emph{Non-volatile Register Emulation} that emulates the persistence of the non-volatile registers in the micro-architecture, and the delays of the read/write operations;
	\item \emph{Energy Consumption Approximation} that approximates the energy consumption of the emulated technology. 
\end{enumerate}
Figure~\ref{fig:system} presents an overview of the proposed architecture. In the following subsections, we summarize the design of the aforementioned auxiliary blocks.

\subsection{Intermittency Emulation}
\label{chap:IE}

This block triggers a \texttt{Reset} signal to emulate a power failure. It can generate random triggers as well as follow an energy trace of a realistic energy harvesting scenario, as presented in \cite{hester2014ekho}.  This block can comprise a memory that can hold a pre-collected voltage trace. Thanks to a prescaler it is possible to choose the frequency at which a new voltage value can be read from this memory. The value can be compared against several pre-determined threshold values using comparators. If the voltage value in the trace is smaller than the corresponding threshold value, the corresponding comparator outputs a high signal. A multiplexer placed in front of the comparators selects the output of the desired threshold comparison operation as the \texttt{Reset} signal.  

\begin{figure}
\centering
	\includegraphics[width=0.6\columnwidth]{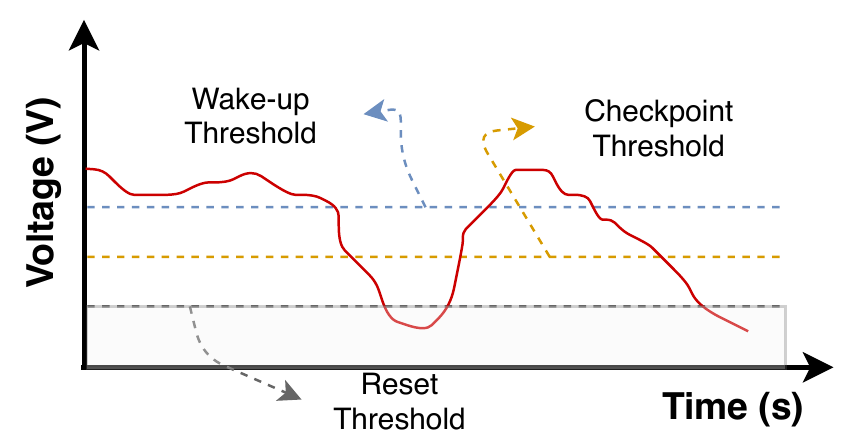}
	\caption{The computing system dies when the voltage level in the capacitor is below the reset threshold. The system starts operating when the voltage level is above the wake-up threshold. The software running on the computing system can also be notified, via the checkpoint threshold, so that the volatile state can be copied manually to the non-volatile memory, as in~\cite{balsamo2016hibernus++}.}
	\label{fig:threshold}
\end{figure}

This block can also output signals that indicate if a threshold condition is satisfied to be used by some other logic in the architecture, as presented in Figure~\ref{fig:threshold}. In particular, a threshold value can be set to trigger a software routine just before a power failure, as in ~\cite{balsamo2016hibernus++}. Preferentially, this block can signal a dedicated hardware module to trigger a backup operation. The signaled hardware block can copy the volatile state elements of the micro-architecture to their non-volatile counterparts.

\subsection{Non-volatile Register Emulation.}
\label{chap:NVRE}

Since the registers in FPGA logic elements are volatile, one cannot directly implement non-volatile registers. Within this block, we implement non-volatile registers using FPGAs' volatile registers via the following design: 
In our emulation architecture, we connected the \texttt{Reset} signal to all volatile registers and combinational logic, i.e., to all components other than the non-volatile registers. Therefore, when the reset signal is triggered, volatile registers are cleared. Since we did not connect the \texttt{Reset} signal to non-volatile register blocks (implemented as volatile registers), their contents will remain in case of resetting other logic elements. Another issue is that non-volatile read/write operations are slower than volatile ones. We added logic that emulates the parametric delays introduced by non-volatile memory circuits. To this end, we placed a logic following the inputs of each volatile register, which emulates non-volatile memory delay.

\subsection{Energy Consumption Approximation}
\label{chap:ECA}

We accelerate the energy estimation by mapping the energy model-related circuit onto the prototyping platform to provide a reliable emulation and assessment of the intermittent micro-architecture.
We defined an energy model for each volatile logic block, as depicted in Figure~\ref{fig:system}. The energy model, fed by activity counters, is configured to measure the energy consumption of the programmed logic into each block. It considers the technology of the emulated chip and the type of activity requested by each volatile logic block. The counters provide in-depth and distributed information about the energy performance. The parametric delays introduced in the read/write NVMs realize the latency of the used memory technology (i.e., ReRAM, FRAM, etc.).

\begin{table}
    \caption{Main parameters characterizing different non-volatile memory technologies.}
    \label{tab:nvmem}
\centering
\begin{tabular}{l|r|r|r|r|r}
Feature / Technology & FeRAM & MRAM & nvSRAM & ReRAM & PRAM \\\hline
Data retention (years ) & 10 \cite{FeRAM} & 20 \cite{MRAM-NXP} & 20 \cite{nvSRAM-datasheet} & 10 \cite{RRAM-Fujutsu} &  10 \cite{PRAM-datasheet} \\
\rowcolor[RGB]{250, 250, 250}
Endurance (cycles) & $10^{15}$ \cite{FeRAM} & $10^8$ \cite{MRAM} & Unlimited \cite{nvSRAM-datasheet} &  $10^6$ \cite{RRAM-Fujutsu} & $10^6$\cite{PRAM-datasheet} \\
Read access time (ns)& 55 \cite{FeRAMDatasheet} & 35 \cite{MRAM-NXP}& 10 \cite{nvSRAM-datasheet-2}&  10 \cite{DBLP:journals/corr/abs-2010-04406} & 115 \cite{PRAM-datasheet} \\
\rowcolor[RGB]{250, 250, 250}
Write access time (ns)& 55 \cite{FeRAMDatasheet} & 35 \cite{MRAM-NXP}& 10 \cite{nvSRAM-datasheet-2}& 50 \cite{DBLP:journals/corr/abs-2010-04406}& 115 \cite{PRAM-datasheet} \\
Sizes (nm)& 130 \cite{FRAM-Wiki}& 14 \cite{MRAM-Wiki}& & 28 \cite{RRAM-Wiki} & 90 \cite{PRAM-Wiki} \\
\rowcolor[RGB]{250, 250, 250}
Read Current (mA) & 8 \cite{FeRAMDatasheet} & 55 \cite{MRAM-NXP} & 3 \cite{nvSRAM-datasheet} & 1.5 \cite{RRAM-Fujutsu} & 30 \cite{PRAM-datasheet}\\
Write Current (mA) & 8 \cite{FeRAMDatasheet} & 105 \cite{MRAM-NXP} & 3 \cite{nvSRAM-datasheet} &   0.15 \cite{RRAM-Fujutsu} & 15 \cite{PRAM-datasheet} \\
\rowcolor[RGB]{250, 250, 250}
Standby Current ($\mu$A) & 90 \cite{FeRAMDatasheet} &18000 \cite{MRAM-NXP} & 250 \cite{nvSRAM-datasheet}  & 60 \cite{RRAM-Fujutsu} & 80 \cite{PRAM-datasheet} \\
Sleep Current ($\mu$A) & 5 \cite{FeRAMDatasheet} & & 8 \cite{nvSRAM-datasheet}  & 6 \cite{RRAM-Fujutsu} &  \\
\rowcolor[RGB]{250, 250, 250}
Read energy \footnote{Energy = Read current $\cdot$ Read Access time $\cdot$ * $V_{dd}$ (=3.3V) } ($pJ$)& 1452 & 6352.5 & 99 & 49.5 & 11385\\
Write energy \footnote{Energy = Read current $\cdot$ Read Access time $\cdot$ * $V_{dd} $ (=3.3V)} ($pJ$)& 1452  & 12127.5 & 99 & 24.75 &  5692.5
\end{tabular}

\end{table}

\sysname implements the non-volatile memory energy consumption model by considering the real-world and already validated parameters provided by the vendors of the non-volatile memories. Table~\ref{tab:nvmem} presents a comparison of five different non-volatile memory technologies based on the the main parameters characterizing them. We considered Ferroelectric RAM (FeRAM), Magnetoresistive RAM (MRAM), Non-Volatile SRAM (nvSRAM), Resistive RAM (ReRAM) and Phase-change RAM (PRAM) technologies, by obtaining parameters from the state of the art commercial chips. The accuracy of \sysname energy approximation (including the timing behavior) depends on the accuracy of these parameters that characterize the selected non-volatile memory technology.

%% file: 4-implementation.tex
\section{\sysname Implementation}
\label{sec:implementation}

After presenting the high-level description of \sysname, we present its implementation details in this section. We implemented \sysname in VHDL using Vivado 2020.1~\cite{vivado}, as a framework to emulate any digital non-volatile logic on FPGAs in the market.
\sysname framework helps designers of transiently-powered systems to test and characterize their architecture by using the provided non-volatile memory abstractions. The implementation of \sysname framework is composed of a set of modules that implements the design presented in Section~\ref{sec:prop_emu_arc}. These modules are:
\begin{enumerate}
    \item Non-volatile Register (NVR), 
    \item Intermittency Emulator (IE), 
    \item Energy Approximator (EA)
    \item Instant Energy Calculator (IEC).
\end{enumerate} 
In the following subsections, we describe the implementation of these modules in detail.

\begin{figure}
    \centering
    \includegraphics[width=0.7\columnwidth]{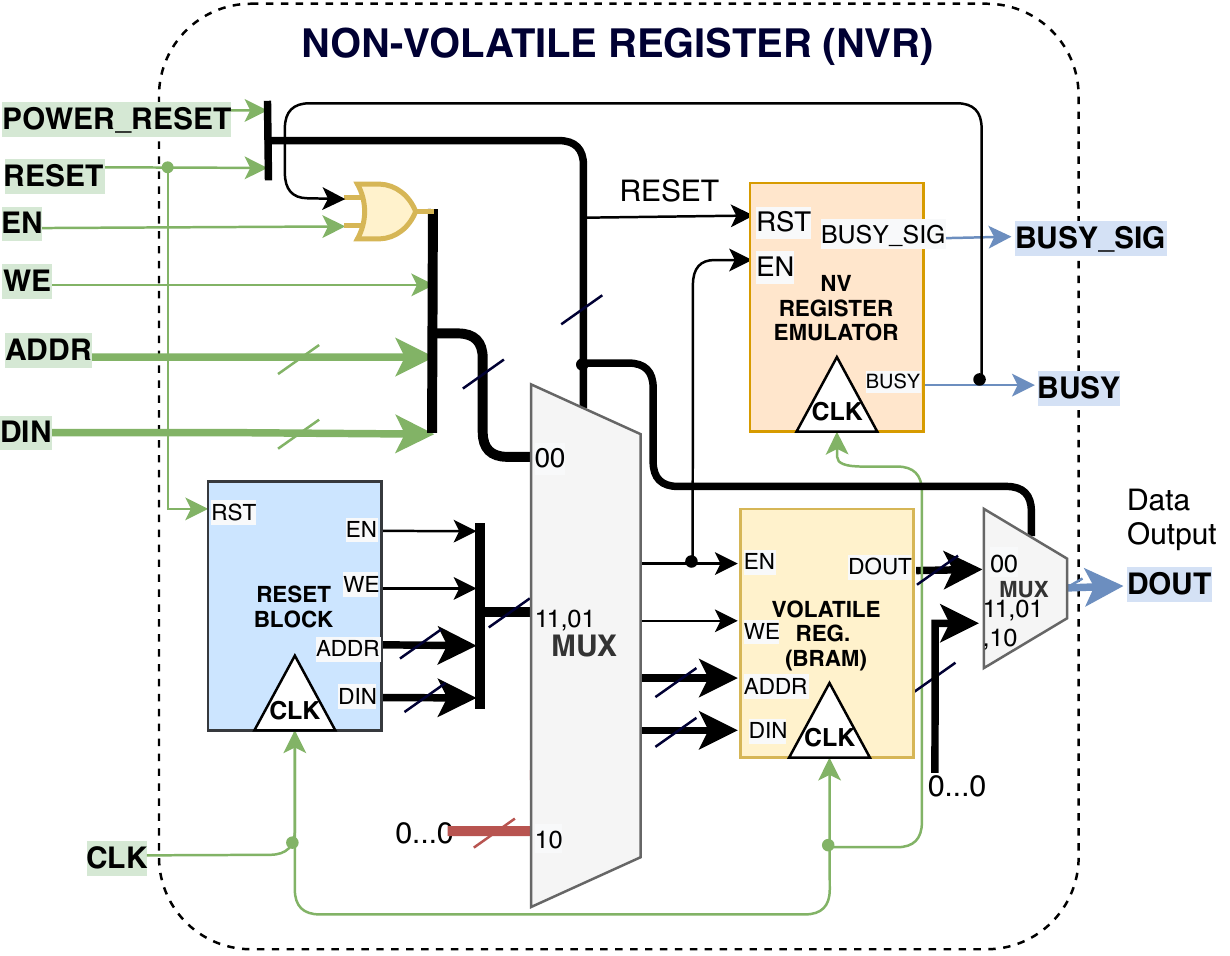}
    \caption{Non-Volatile Register (NVR) block diagram. This entity groups: non-volatile memory emulator (NVRE), volatile register (BRAM), volatile memory cleaner (RESET BLOCK), and the multiplexer that imposes the right input signals to the volatile memory. 
    \signal{RESET} and \signal{POWER\_RESET} signals are the control inputs of the memory multiplexer and output multiplexer. Memory multiplexer selects signals for the input ports of the volatile register while the output multiplexer imposes either the volatile register output or zeros as the data output of NVR. The zero condition is met for both multiplexers only when \signal{POWER\_RESET} is on and \signal{RESET} off.}
    
    \label{fig:nvr_blk_diag}
\end{figure}

\subsection{Non-Volatile Register (NVR)}
\label{subsec:nvr}
The main blocks, input/output signals, and internal connections of the non-volatile register (NVR) are presented in Figure~\ref{fig:nvr_blk_diag}. We implemented the memory that holds NVR data using Xilinx block memory (BRAM) proprietary IP ~\cite{xilinxbramipdoc}, but other open source BRAM implementations can also be used. Inputs to a BRAM block are address bus (\signal{ADDR}),  input data (\signal{DIN}), clock (\signal{CLK}), write enable (\signal{WE}) and enable (\signal{EN}). These signals are also the general inputs of the NVR, as depicted in Figure~\ref{fig:nvr_blk_diag}. Apart from these signals, \texttt{RESET} is the internal hardware reset of the FPGA, and \texttt{POWER\_RESET} is the reset signal that emulates the power failure (from intermittency emulator block, which we will present later). It is worth mentioning that \texttt{POWER\_RESET} does not erase the volatile memory of the block. NVR also holds a non-volatile register emulator (NVRE) block and a reset block (RB). 

\subsubsection{Non-volatile Register Emulator (NVRE)} 

Without the non-volatile register emulator (NVRE) block, NVR behaves like an ordinary volatile memory/register. NVRE imposes a strict access policy that takes into account non-volatile memory is slower than volatile one. This entity is instantiated by providing a time delay (expressed in nanoseconds), which defines the access delay of the emulated non-volatile memory. Hence, this component expects a scaled access time concerning system clock speed. The access time (i.e., the delays due to read and write operations) can easily be obtained from the data sheets of non-volatile memory components, as depicted in Table~\ref{tab:nvmem}. It is also worth mentioning that the aging of the non-volatile memories is not a concern for our framework since the new memory technologies have a high write endurance. As an example, FeRAM  has $10^{15}$ write endurance. Therefore,  even 150000 write operations per second will lead to almost 211 years lifetime. 

The delayed access time is enabled by a busy signal that informs endpoints about the operational status of the component. NVRE has three input signals, i.e., clock (\signal{CLK}), reset (\signal{RST}) and enable (\signal{EN}). NVRE implements the following emulation protocol:
\begin{enumerate}
    \item The input signal \signal{EN} is used to enable NVR access. This signal is also connected to the main BRAM block that holds the NVR data. \signal{EN} is captured on the rising edge of the clock.
    \item Once the NVR is accessed and \signal{EN} is asserted, output signal \signal{BUSY} is also asserted. This signal stays high for a period of the non-volatile access delay. During this period, all the memory-related input ports of the non-volatile register \textbf{must} be kept constant. 
    \item The output data of the NVR (represented by the \signal{DOUT} output of NVR) can be captured when \signal{BUSY} is low.
\end{enumerate}
NVRE implements a counter to count down from the time delay to trigger the \signal{BUSY} signal. NVRE entity also outputs an extra signal (i.e., the \signal{BUSY\_SIG} signal), which is similar to \signal{BUSY}, but pulled low one clock cycle before \signal{BUSY} is pulled low. This signal can be used by synchronous processes to update the input ports of the NVR, in order not to demand extra clock cycles and operate continuously while \signal{BUSY} is primarily intended for asynchronous circuitry.

\subsubsection{Reset Block (RB)}

This entity fills the BRAM with zeros while the whole FPGA resets, i.e., the user pressed the hardware reset button. The \signal{RESET} signal is asserted, which enables the reset block. The reason behind this module is that volatile register (BRAM block) contain old data after a real reset because the FPGA is not powered off. Hence this block wipes all memory to a initial defaults state. To achieve a complete memory wipe the \signal{RESET} signal should be high for a number of clock cycles that equals the size of the non-volatile register to clear the whole BRAM. 

\subsubsection{Operation Consistency of NVR}

The \signal{POWER\_RESET} can be triggered in the middle of an ongoing write operation to the NVR. In reality, a reset during a write operation to non-volatile memory (like Fe-RAM) does not leave non-volatile memory partially updated, i.e., either the word is written or not. In \sysname, we followed a similar strategy to mitigate the side effects of power failures during NVR write operations. \sysname guarantees that if the write operation is accepted by NVR (\signal{BUSY} on), then the operation will be completed successfully (meaning that data is written).

\begin{figure}
    \centering
    \includegraphics[width=0.7\columnwidth]{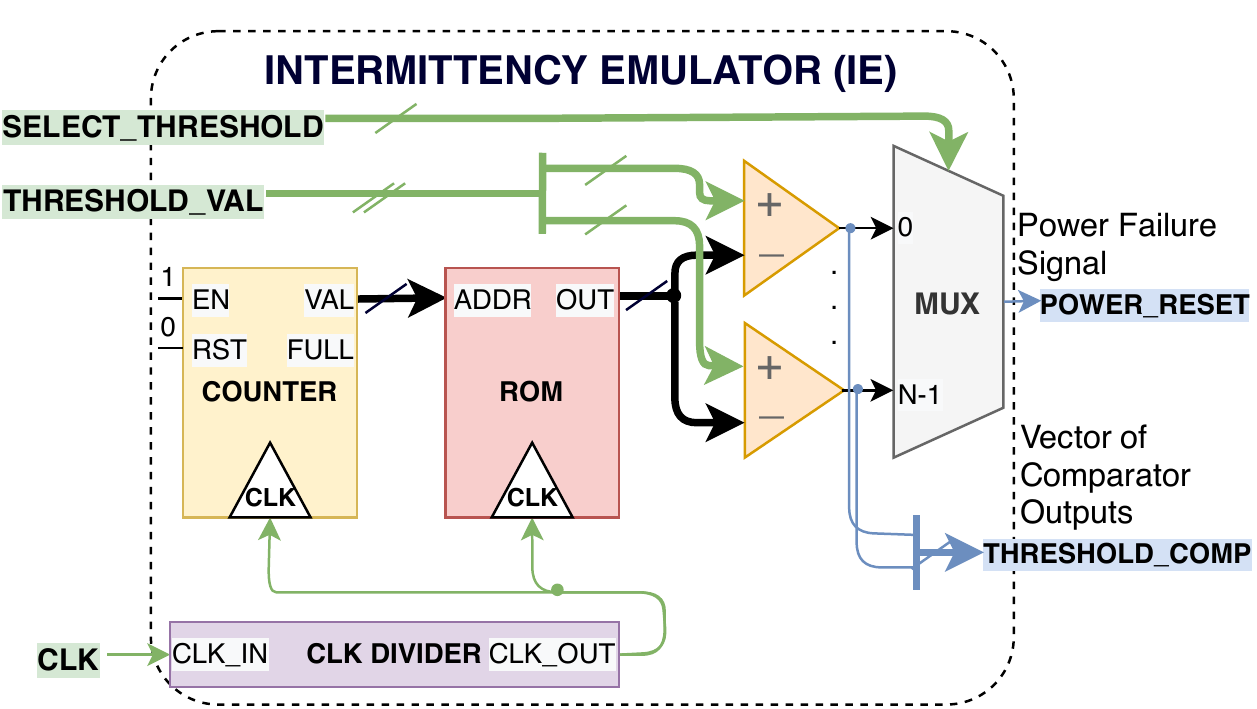}
    \caption{Intermittency Emulator Entity. This entity generates power failure signals considering the selected threshold and voltage values stored in ROM.}
    \label{fig:ie_diag}
\end{figure}

\subsection{Intermittency Emulator (IE)}
\label{sec:intermittency-emulator}

Intermittency Emulator (IE), depicted in Figure~\ref{fig:ie_diag}, implements the auxiliary module presented in Section~\ref{chap:IE}. This entity comprises a ROM memory to hold a voltage trace and a counter that iterates through the entries of the ROM. IE can also be prescaled (via the CLK DIVIDER block in Figure~\ref{fig:ie_diag}) to change trace duration (i.e., to slow down the iteration speed). Once the trace ends (the counter overflows), it restarts from the beginning. Inputs to this entity are \signal{THRESHOLD\_VAL} that is the set of desired voltage thresholds which will be compared against the current value in the ROM memory cell, and \signal{SELECT\_THRESHOLD} which is used to select the comparator whose output generates the desired \signal{POWER\_RESET} signal. Even at the synthesis level, IE can be configured to have multiple comparators allowing having more precise and granular control on the voltage status during runtime. Each comparator checks if the entry of the voltage trace (pointed by the counter) is below a given threshold value, every threshold is provided in advance by the user that can define both the quantity and the values. The multiplexer selects which of the provided thresholds is the one that triggers the \signal{POWER\_RESET}, the one that will be used by all entities of the volatile architecture as the main signal that resets the system due to a power failure. This entity also outputs \signal{THRESHOLD\_COMP}, which is a vector in which each bit indicates if voltage value is higher or lower than the corresponding threshold.

\subsection{Energy Approximator (EA)}

Energy Approximator (EA) entity is composed of a set of counters that are incremented by one at each clock cycle. The number of entities whose energy needs to be approximated determines the number of counters, which can be configured in the source code. Each counter expresses the energy consumption of an entity (during the time it is on) in terms of the clock cycles. The Instant Energy Calculator (IEC), explained shortly, computes the approximated energy consumption of the entities based on the number of clock cycles kept on the counter. It is worth mentioning that the more precise the parameters in Table~\ref{tab:nvmem} are, the more accurate the energy approximation is.

\begin{figure}
    \centering
        \includegraphics[width=0.7\columnwidth]{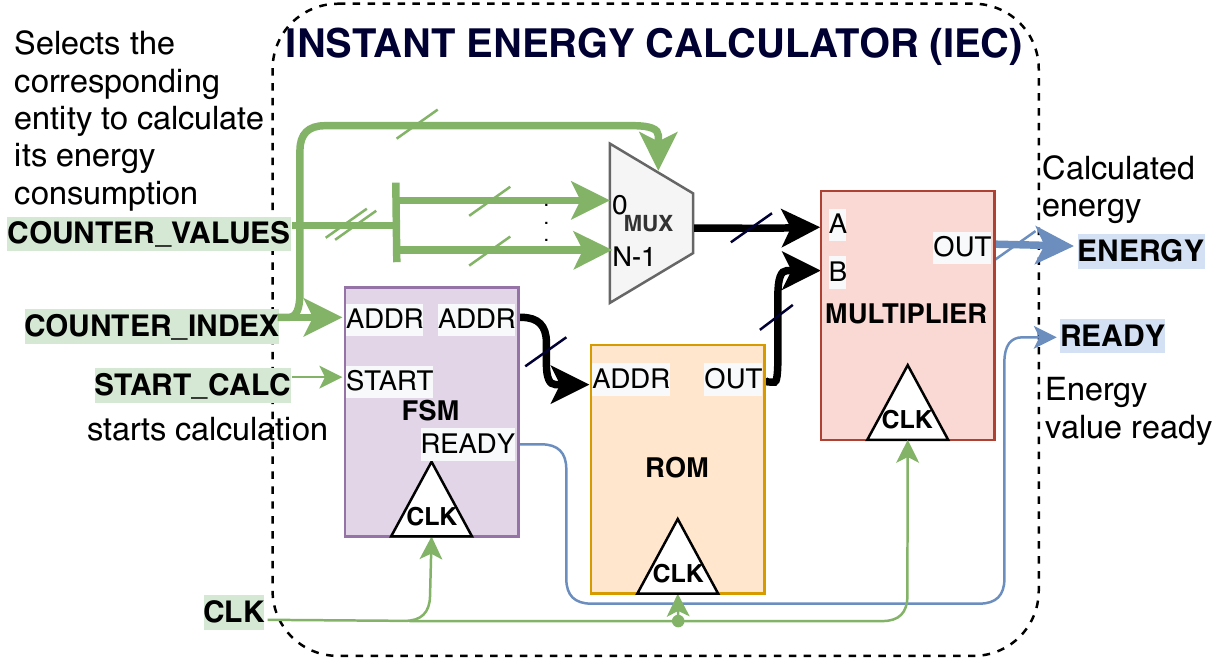}
    \caption{Instant Energy Calculator block diagram. IEC is composed of a ROM that stores Energy Consumption per Clock Cycle (E3C) for each counter in EA, a multiplier that computes the product, and a finite state machine that regulates the operation.}
    \label{fig:IEC_blk_diag}
\end{figure}

\subsection{Instant Energy Calculator (IEC)}

Instant Energy Calculator (IEC), whose implementation is depicted in Figure~\ref{fig:IEC_blk_diag}, converts the number of clock cycles held in the counters of EA into an energy value. IEC takes \signal{EA\_VALUES\_ARRAY} that holds all values of counters in EA and \signal{INDEX} the offset for the previously mentioned array as an input to calculate the energy consumption of the corresponding entity. The calculation is triggered by the \signal{START\_CALC} signal. The energy consumption (per clock cycle) of all components is stored in ROM and defined by the user. It is worth mentioning that the energy consumption of the components can be obtained from their data sheets as well as by performing testbed measurements to observe the actual energy requirements. An internal finite state machine (FSM) manages all operations, like reading values from ROM and multiplication. An important point worth mentioning is that if a process runs long enough, the counters of EA can overflow. Consequently, larger registers and a bigger multiplier are required to hold the number of clock cycles and perform the calculation of the energy consumption. To remedy this issue, IEC 
calculates the approximated energy within a time interval by sampling EA counters at regular intervals and re-initializing the counters in EA. The outputs of IEC are \signal{ENERGY} that holds the calculated energy and \signal{EVALUATION\_READY} that indicates that the calculation is finished. The \signal{ENERGY} output can be accumulated in a shared memory location (e.g., DRAM) to further add up and process the approximated energy consumption.

%% file: 5-evaluation.tex
\begin{figure*}
\centering
	\includegraphics[width=\textwidth]{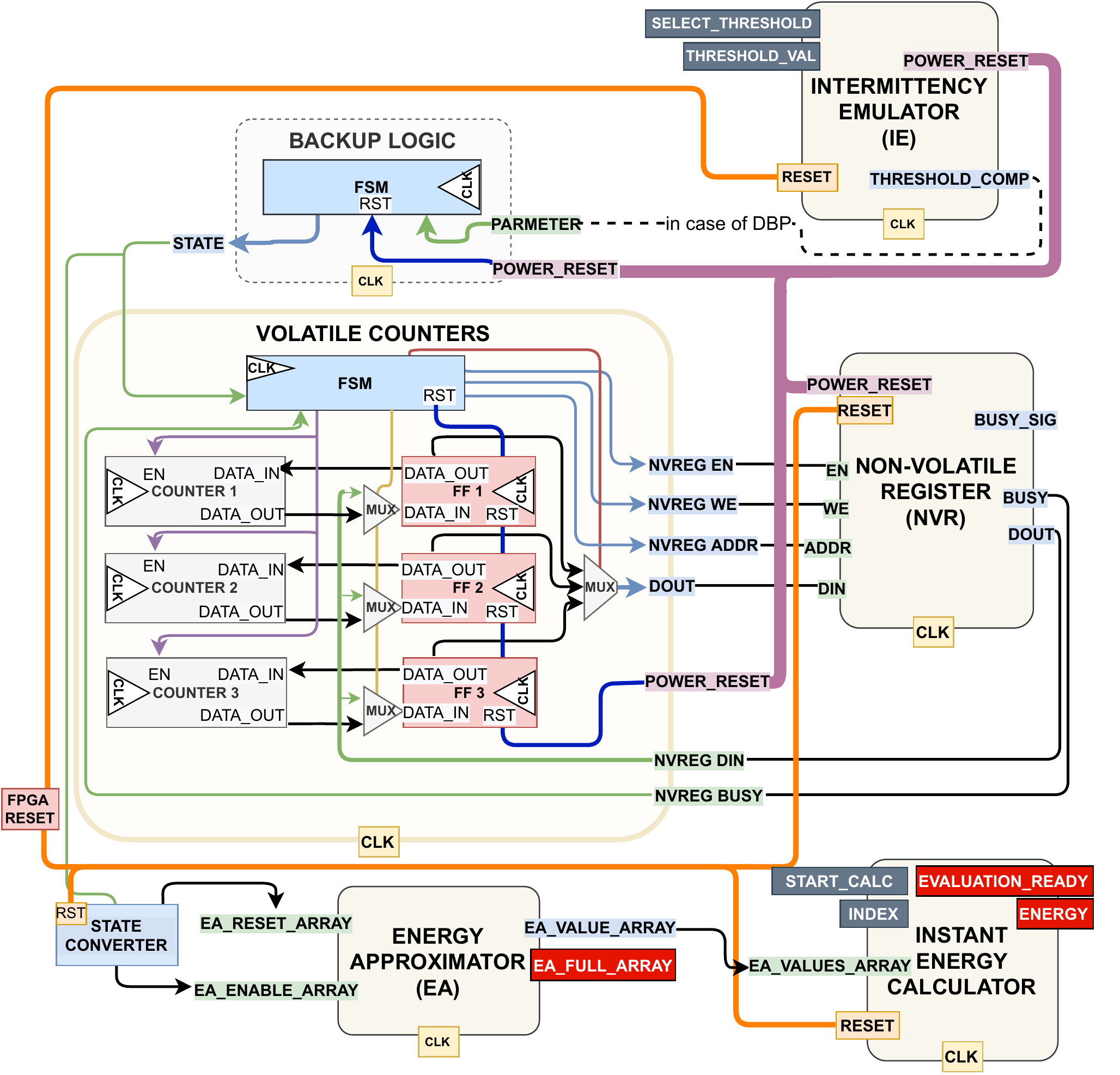}
	\caption{The top-level view of the simulation architecture (together with \sysname blocks). Gray input signals (\signal{SELECT\_THRESHOLD}, \signal{THRESHOLD\_VAL}, \signal{START\_CALC} and \signal{INDEX}) should be given and red output signals (\signal{ENERGY}, \signal{EA\_FULL\_ARRAY} and \signal{EVALUATION\_READY}) should be captured by the end user.} 
	\label{fig:top_level_view}
\end{figure*}

\begin{figure}
    \centering
	\includegraphics[width=0.5\columnwidth]{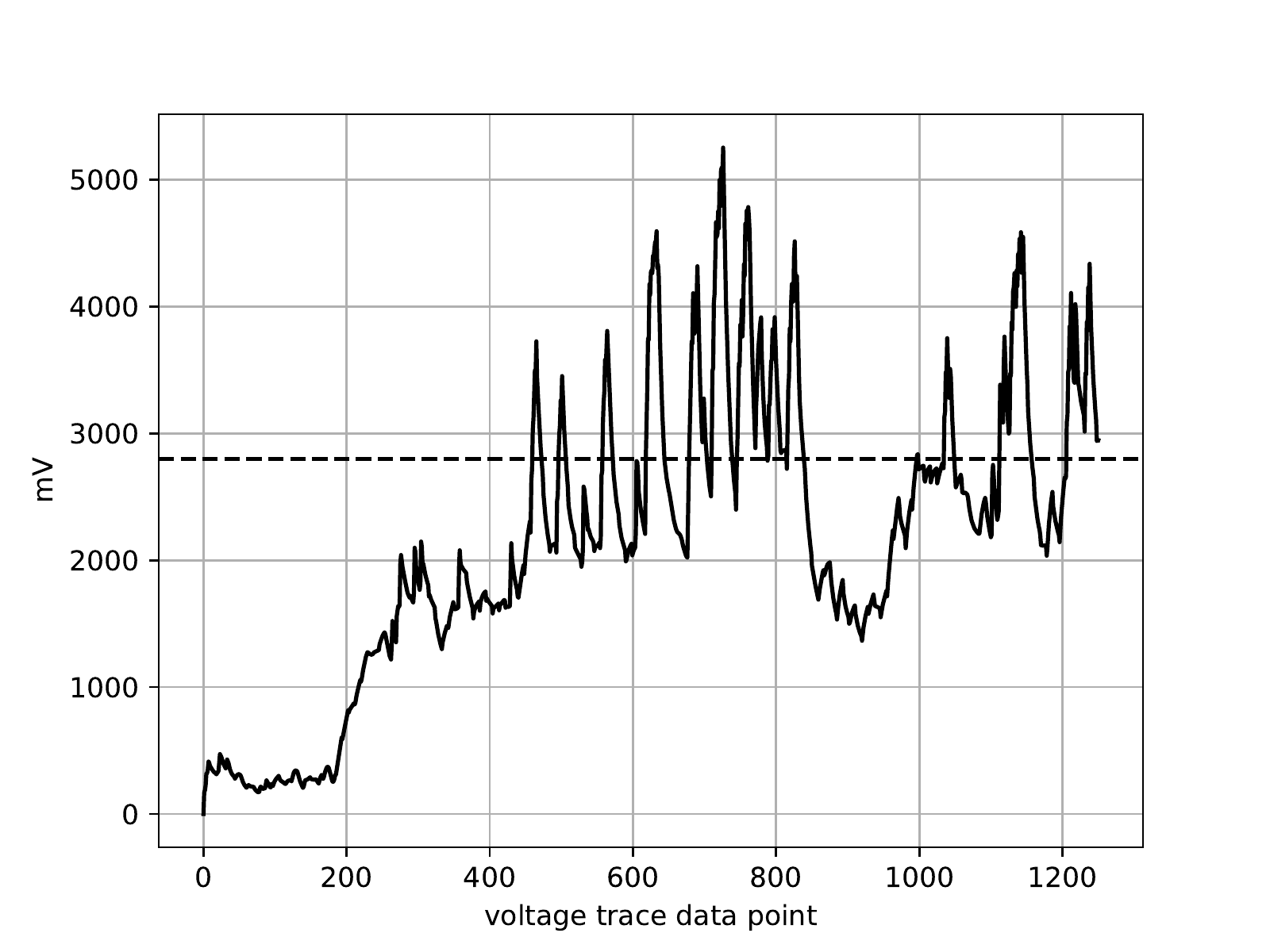}
	\caption{Voltage trace used in simulations---taken from~\cite{persistlab}. This voltage trace depicts the voltage level of a capacitor  that stores the harvested energy from an RFID reader.  We averaged this trace in groups of 25 samples to reduce memory usage. The dotted line is the threshold used to simulate a shutdown event (\texttt{POWER\_RESET}), which we set to 2.8\,V. The total shutdown time is 75\% of the trace.} 
	\label{fig:v_trace}
\end{figure}

\begin{figure*}
     \centering
     \begin{subfigure}[b]{0.32\textwidth}
         \centering
         \includegraphics[width=\textwidth]{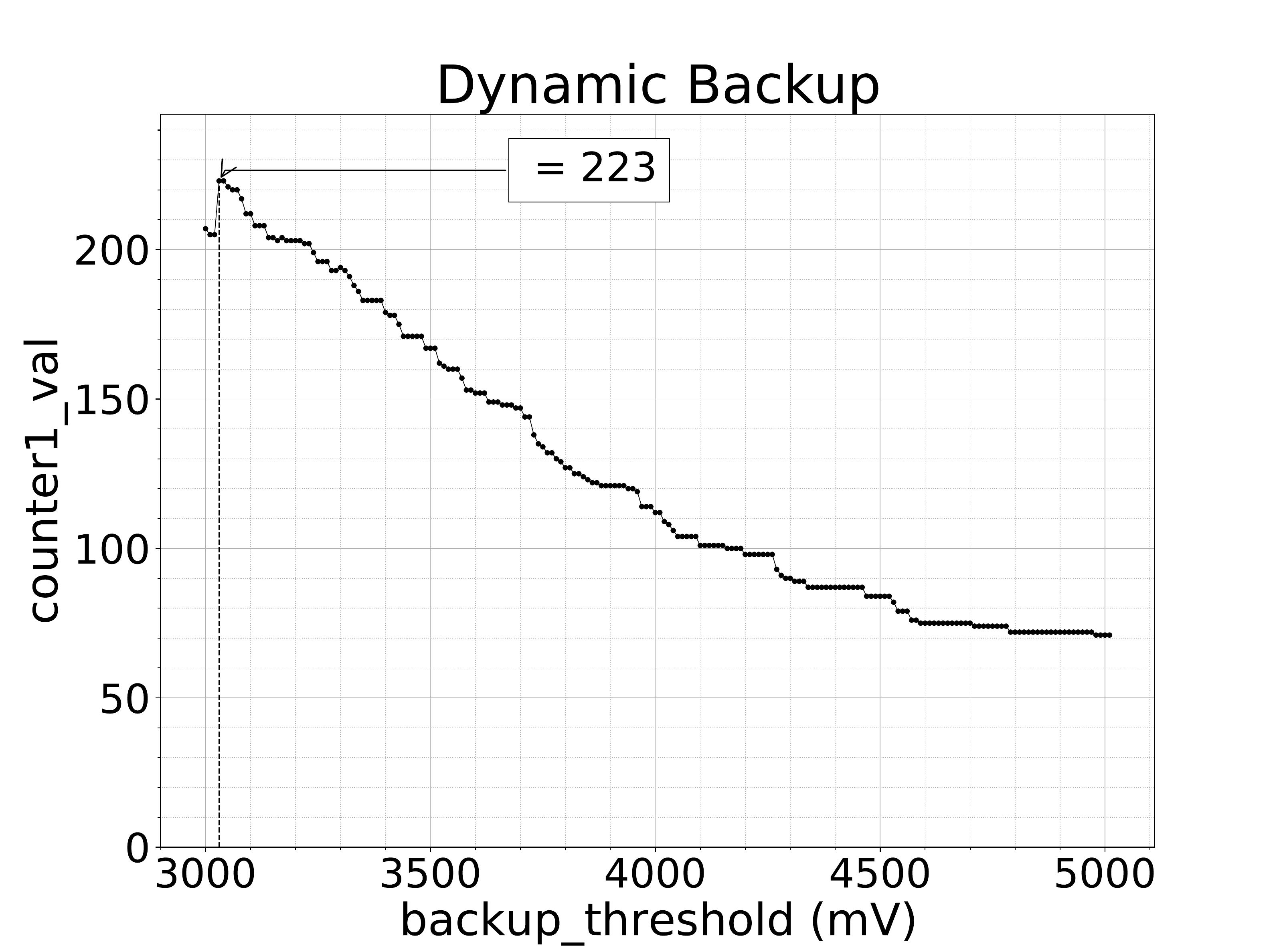}
     \end{subfigure}
     \hfill
     \begin{subfigure}[b]{0.32\textwidth}
         \centering
         \includegraphics[width=\textwidth]{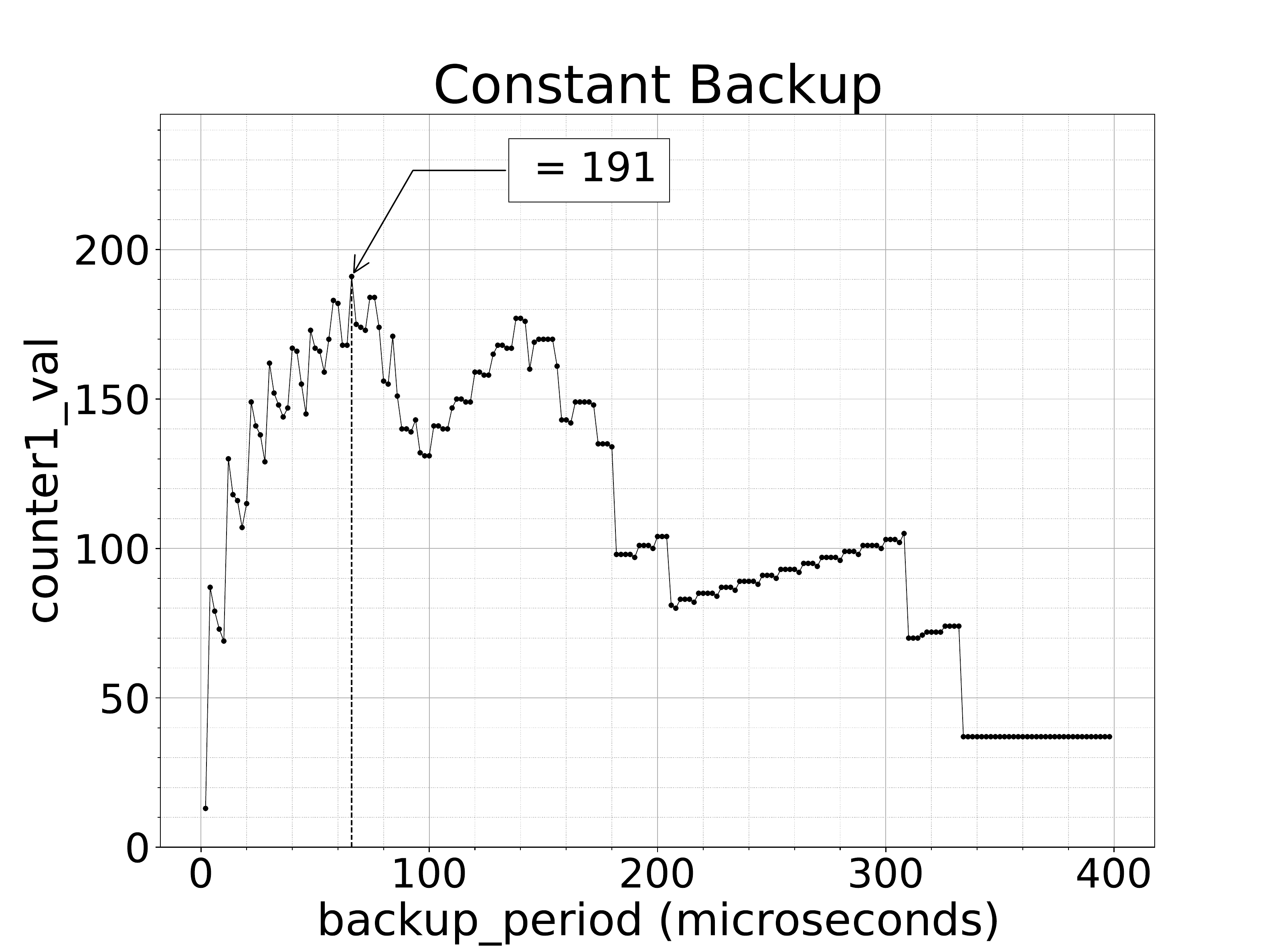}
     \end{subfigure}
     \hfill
     \begin{subfigure}[b]{0.32\textwidth}
         \centering
        \includegraphics[width=\textwidth]{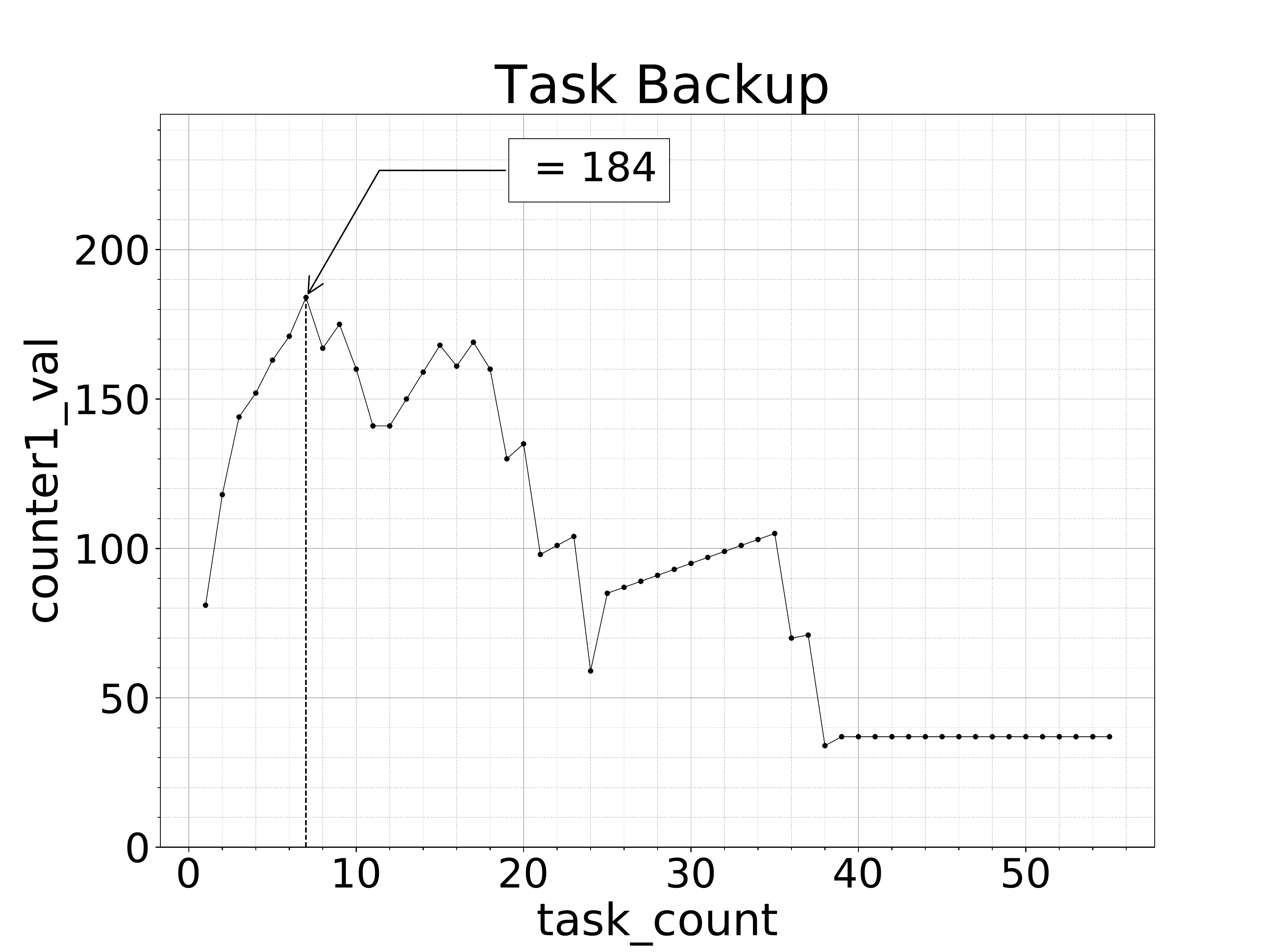}
     \end{subfigure}
        \caption{Comparison of counter 1 value, among the different policies.}
    \label{fig:results_cntr1_val}
\end{figure*}

\begin{figure*}
     \centering
     \begin{subfigure}[b]{0.32\textwidth}
         \centering
         \includegraphics[width=\textwidth]{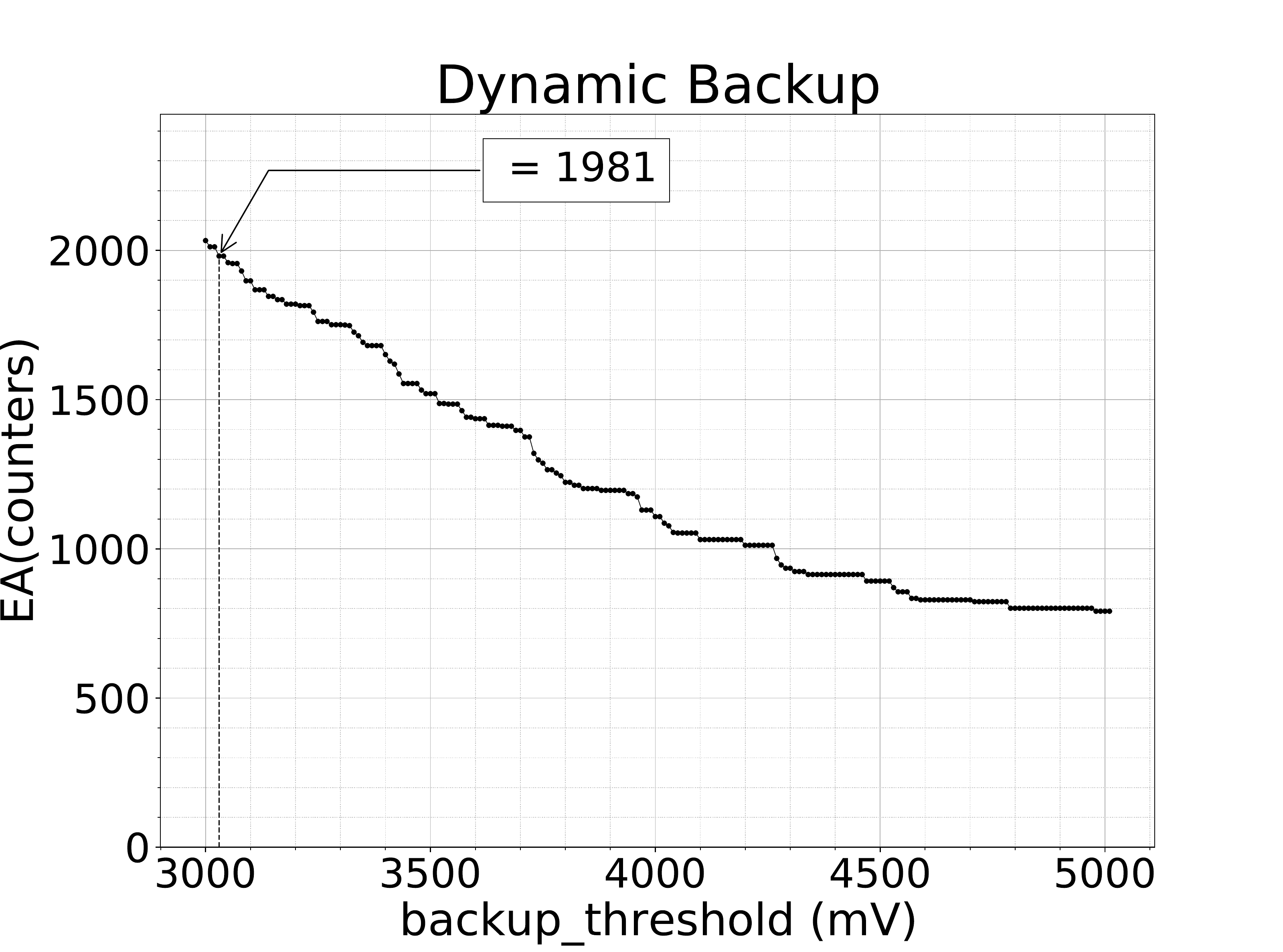}
     \end{subfigure}
     \hfill
     \begin{subfigure}[b]{0.32\textwidth}
         \centering
        \includegraphics[width=\textwidth]{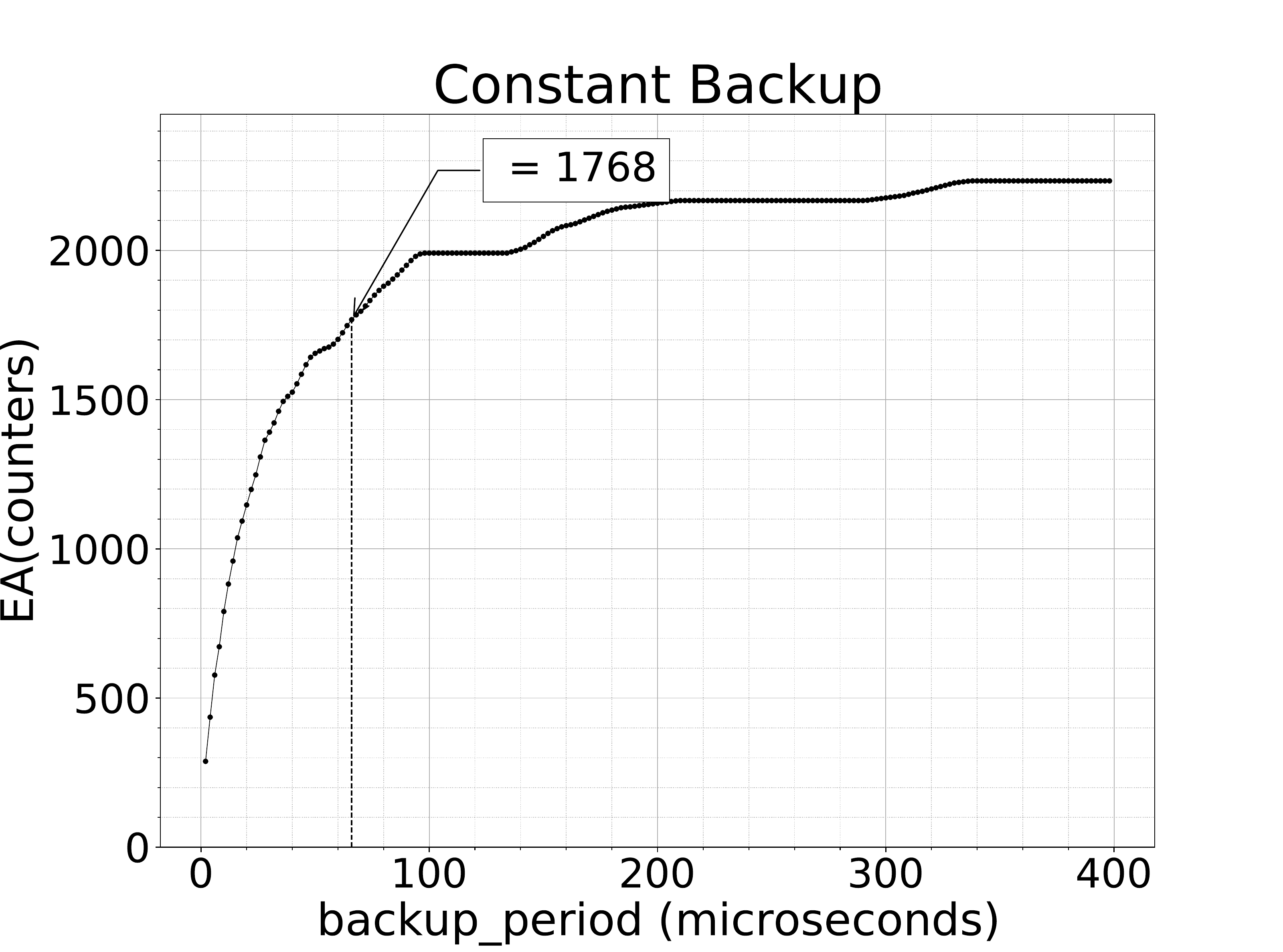}
     \end{subfigure}
     \hfill
     \begin{subfigure}[b]{0.32\textwidth}
         \centering
        \includegraphics[width=\textwidth]{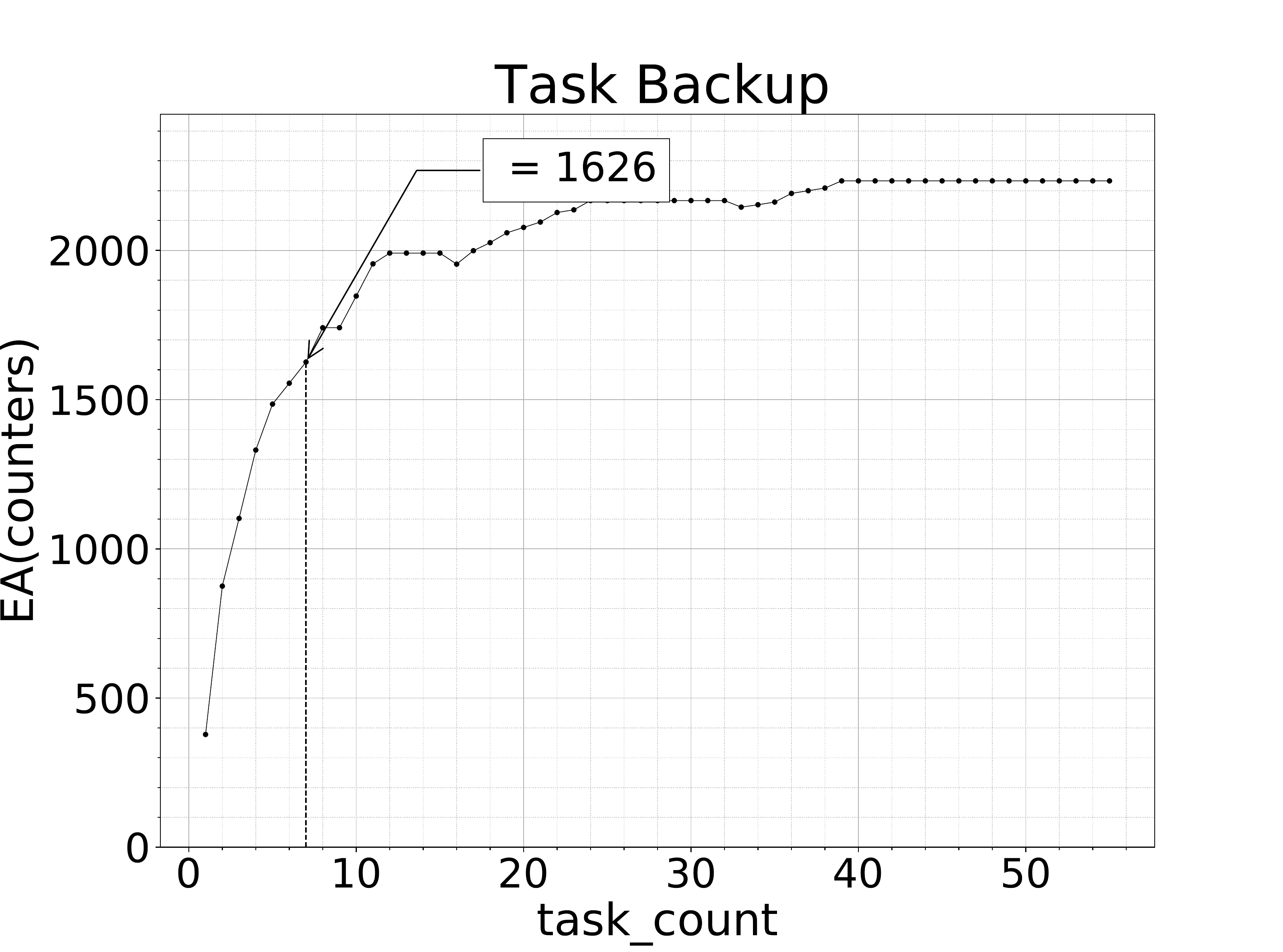}
     \end{subfigure}
     \caption{Comparison of the approximated energy consumption of all counters using the different policies.}
    \label{fig:results_ea_counters}
\end{figure*}

\begin{figure*}
     \centering
     \begin{subfigure}[b]{0.32\textwidth}
         \centering
          \includegraphics[width=\textwidth]{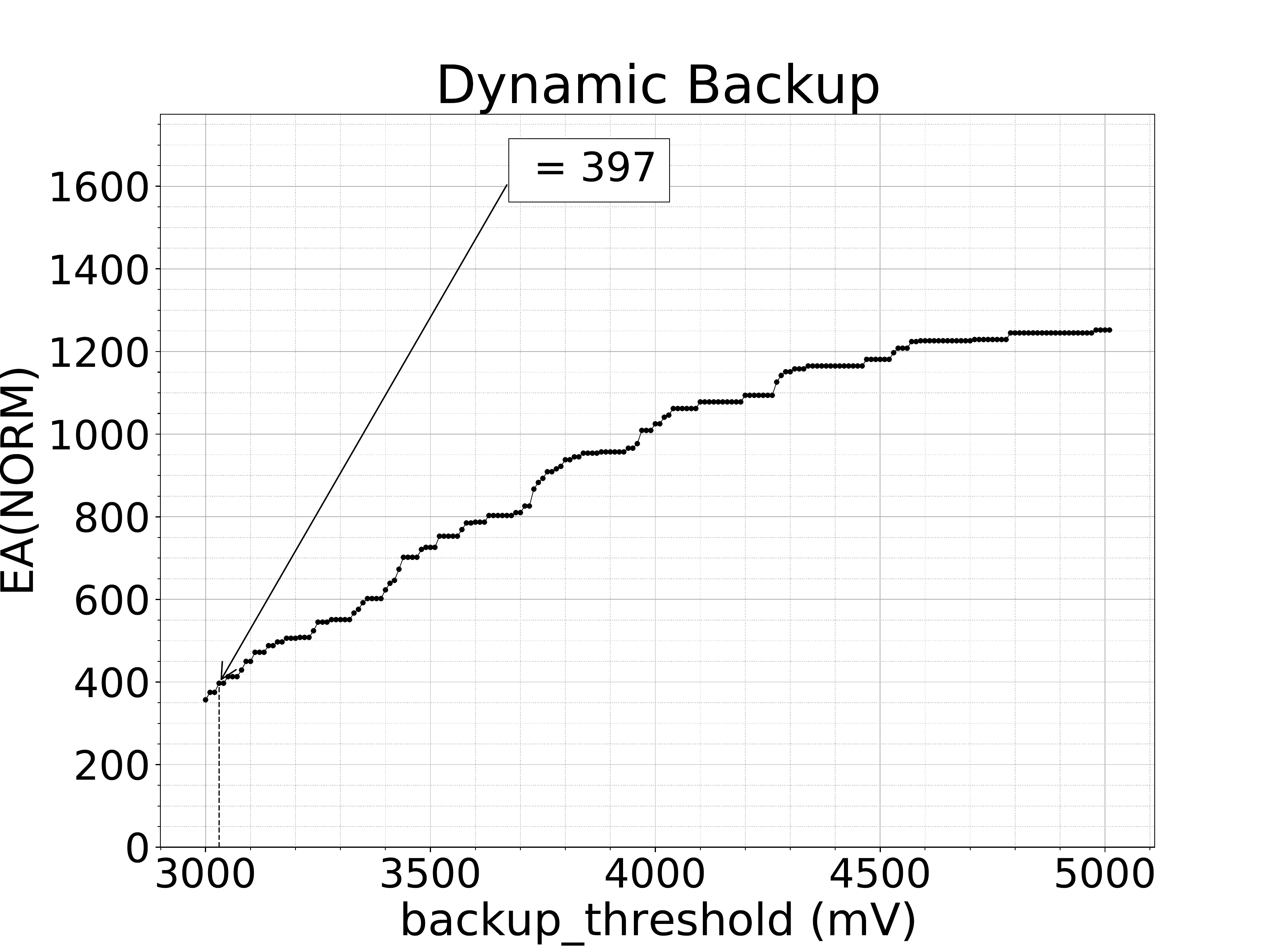}
         \label{fig:y equals x}
     \end{subfigure}
     \hfill
     \begin{subfigure}[b]{0.32\textwidth}
         \centering
         \includegraphics[width=\textwidth]{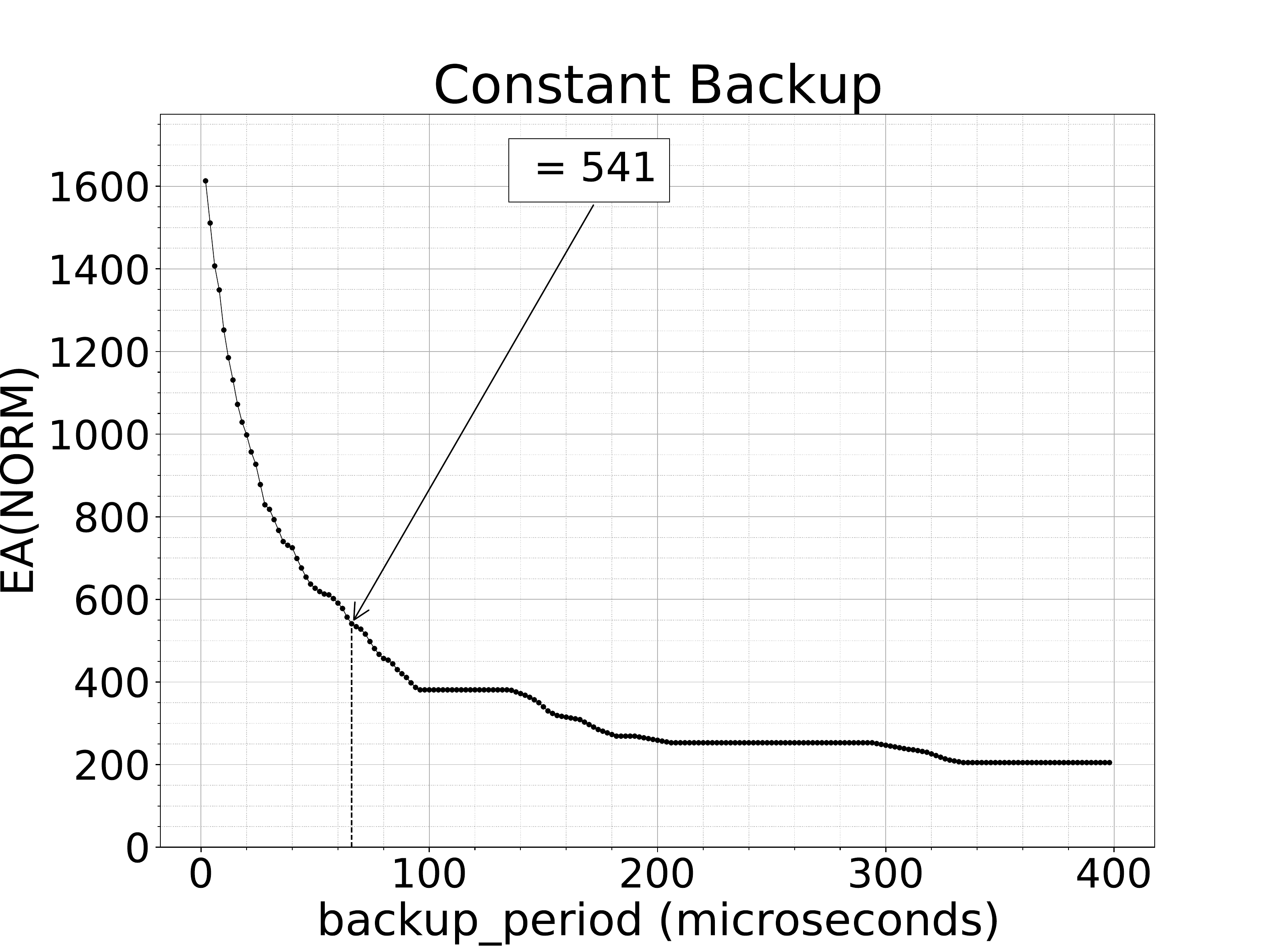}
         \label{fig:three sin x}
     \end{subfigure}
     \hfill
     \begin{subfigure}[b]{0.32\textwidth}
         \centering
        \includegraphics[width=\textwidth]{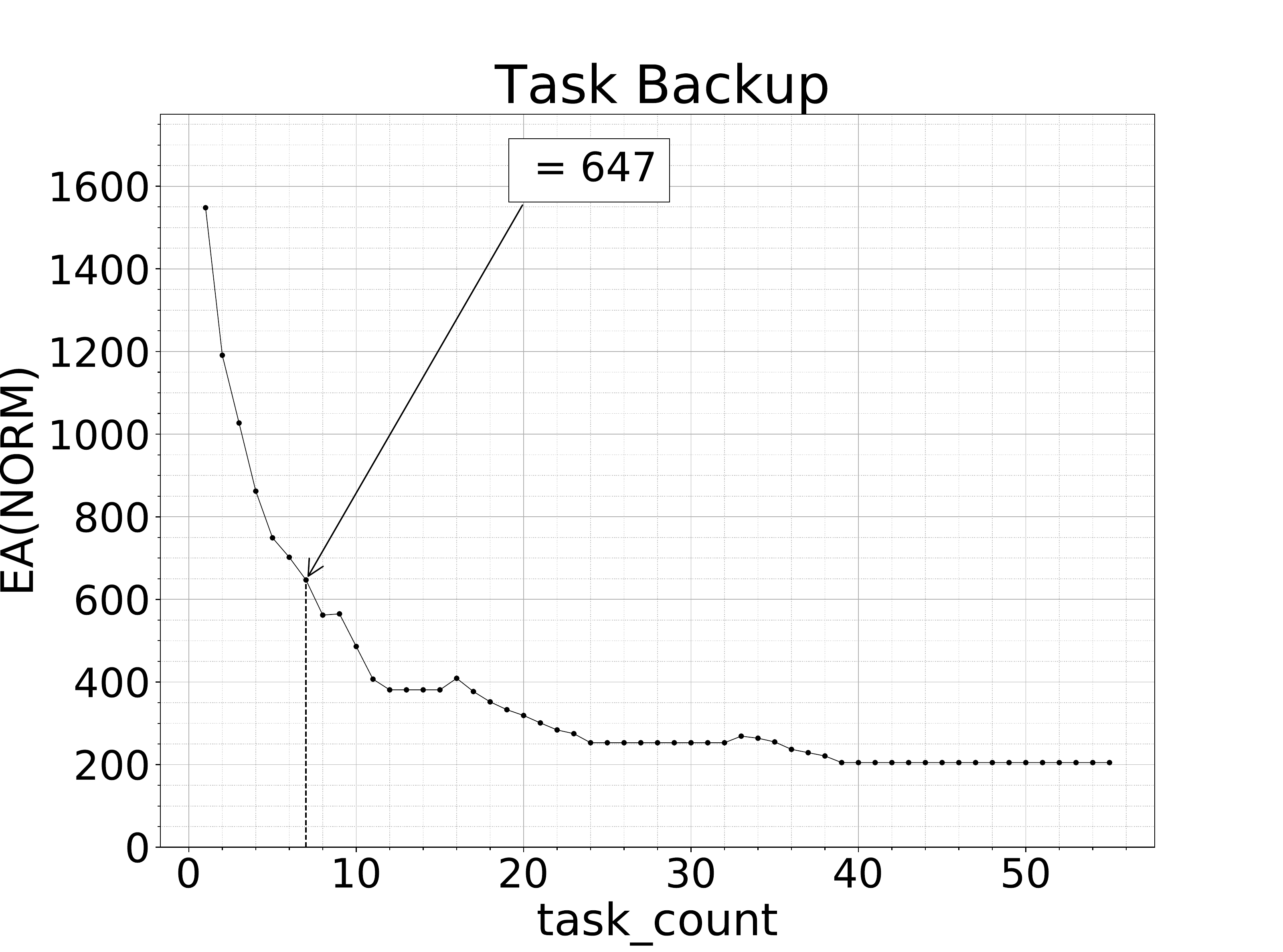}
         \label{fig:five over x}
     \end{subfigure}
     \caption{Comparison of the approximated energy consumption of \sysname using the different policies.}
    \label{fig:results_ea_norm}
\end{figure*}

\section{Evaluation of \sysname}
\label{sec:evaluation}

In this section, we present our simulations, performed via Vivado Simulator~\cite{simulator}, to understand how \sysname can be leveraged to emulate a custom non-volatile logic (which keeps its state upon reset) together with a volatile logic (which loses its state upon reset). The details of the architecture implemented for our simulations and evaluation are given as follows.

\subsection{Simulation Architecture} 
\label{sec:simulation-architecture}

We implemented an architecture that comprises a series of three counters (which lose their values upon power failures), and a backup logic that implements a backup policy which regulates how frequently the system gets its state stored in non-volatile memory (a backup). This operation needs access to non-volatile registers (described in Section~\ref{subsec:nvr}) hence the backup-logic regulates these transactions. The overall blocks forming the simulation architecture (together with \sysname) and their connections are presented in Figure~\ref{fig:top_level_view}. Multiple backup policies can be implemented as different finite-state-machines in the backup logic block. Each backup policy leads to different simulation results of the emulated transiently-powered system since it changes the run-time behaviour. With the exception of the non-volatile registers, all entities of the simulation architecture are thought as volatile, hence after an emulated power off, the components lose state. Simulation architecture implements store/recover operations of the counter registers into/from the non-volatile registers. Backup logic triggers these operations. Therefore, counters continue counting from where they left after an emulated power failure. 

\subsubsection{Volatile Counters}
As depicted in Figure~\ref{fig:top_level_view}, the Volatile Counters block comprises a finite state machine (denoted as FSM), three array of flip-flops to hold counter values (denoted as FF), and three counter blocks that increment the values stored in the corresponding array of flip-flops (denoted as COUNTER). During normal operation each counter is increased sequentially and with different base values. Specifically, during the increment operation,  the counter value is fetched from the volatile memory first (i.e., the corresponding flip-flop array). Then, this value is incremented, and the result is saved in the same flip-flop array. Finally, the FSM block selects the next counter, and this operation is performed again, and so on. As mentioned previously, \sysname emulates a power failure by setting the \signal{POWER\_RESET} signal. Since the FPGA is still on during this operation, the volatile counters do not lose their state. Therefore, FSM emulates a real memory reset process by clearing the dedicated array of flip-flops of the counters.

\subsubsection{Backup Logic}
The backup logic implements a finite state machine that calls store/recovery procedures that access the non-volatile register (NVR). The order and rate of these procedures define the backup policy implemented by the finite state machine. More precisely the logic can be tuned by means of an external parameter. We implemented three backup policies for our simulations:

\begin{enumerate}
    \item {\bf Dynamic Backup Policy (DBP):} DBP uses the output of \ie as a dynamic input parameter, i.e. the state of an internal comparator bounded to a a voltage threshold (see comparators in \ref{sec:intermittency-emulator}). Hence the tuning parameter for DBP is the set of input voltage thresholds in \ie. Thresholds are defined in advance by considering the characteristics of the emulated transiently-powered architecture {and voltage trace}. For simplicity, we used only one threshold to trigger the backup operation in our current DBP implementation. When the current voltage trace value drops below the given \emph{backup threshold} value, DBP goes into the hazard mode to save the state of the volatile counters into NVR.
    \item {\bf Constant-time Backup Policy (CBP):} CBP  uses a user-defined \emph{backup period} to backup the counters periodically. This constant is the tuning parameter for this backup policy. A timer initialized with this value triggers the backup when the time runs out, then the cycle restarts. As in DBP the possible values the parameter can assume are limited by the architecture and the voltage trace. 
    \item {\bf Task-based Backup Policy (TBP):} TBP backups the system on predefined computation boundaries. The policy tuning parameter is called  \emph{backup task count}  and defines the necessary goal that the \emph{simulation architecture} must reach before a backup can be issued. The time interval spanned by the voltage trace should be long enough to have at least one backup operation.
\end{enumerate}
It is worth mentioning that the registers forming the finite state machine implemented by the Backup Logic are volatile and lose status after an emulated power failure. Therefore, the first operation that all finite state machines perform after a shutdown is the recovery procedure, this lets counters restore their values. 

\subsection{Simulation Results}

We performed simulations concerning different backup policies to validate the architecture described in section~\ref{sec:simulation-architecture}. For each backup policy, we performed multiple runs of simulations with different parameter values, to understand the effect of these backup policies. In all simulation runs, we used the voltage trace depicted in Figure~\ref{fig:v_trace}, which spans 100 microseconds with a system clock of 100 MHz. We set the access/request time of the NVR in \sysname to 80 nanoseconds, which means that every process must wait for at least eight clock ticks to perform another request from the non-volatile memory. This value is compatible with similar non-volatile technologies like FeRAM~\cite{FRAM}.

\subsubsection{Metrics.} For each simulation run, we collected the following values:

\begin{enumerate}
    \item {\bf Counter Value:} This metric represents the value inside the counter 1 (depicted in Figure~\ref{fig:top_level_view}). Since the FSM increases the counters one at a time, each counter should wait that other counters get updated before increasing. Therefore, counter 1 is incremented with a frequency of 4.16\,MHz under normal execution.
    \item {\bf Energy Approximation (Counters):} This metric represents the approximated energy consumption (calculated by the EA) of all volatile counters depicted in Figure~\ref{fig:top_level_view}.
    \item {\bf Energy Approximation (\sysname):} This metric represents the approximated energy consumption of all components forming \sysname. 
\end{enumerate}

\subsubsection{Parameters}
As mentioned previously, backup policies have different parameters, i.e., \emph{backup threshold}, \emph{backup period} and \emph{backup task count}, respectively. In our simulations, we selected different values for these parameters. More precisely:
\begin{itemize}
    \item {\bf DBP backup threshold (mV):} This value defines the point below which the architecture must stop and start performing a backup. It is compared against the voltage level of \ie. While the voltage level is below the threshold, computation will not progress. The parameter starts from 3000 mV and is incremented with a step size of 10 mV until 5010 mV is reached. 
    \item {\bf CBP backup period (microseconds):} This value defines a period used by an internal timer. When the timer fires, the architecture must perform a backup if there is still available energy. This parameter starts with 2 microseconds and is incremented with a step size of $2us$ until 398 microseconds are reached.
    \item {\bf TBP backup task count (numerical value):} This value defines a target goal that \signal{COUNTER1} must reach before the architecture can perform a backup. More precisely, when the counter value is a multiple of this value, the architecture stops and performs a backup. The range of possible values for \textit{backup task count}  is between 1 to 55.
\end{itemize}

\subsubsection{Results}
We run our simulations multiple times for each parameter value and using the same voltage trace and duration. Figure~\ref{fig:results_cntr1_val} presents the value of COUNTER 1 (\signal{counter1\_val}) considering three policies. This metric shows the amount of progress established (i.e., counter increment operation)  despite the emulated power failures. Therefore, the greater the value of this metric is, the superior the backup policy is. During simulations, DBP exhibited the best case when the \emph{backup threshold} equals 3040\,mV. This is since DBP carries out backups only when necessary (following the voltage trend), without wasting time and execution. Moreover, the higher the backup threshold is, the lesser the available time for computation will be. Therefore, the \signal{counter1\_val} metric decreases with respect to the increased \emph{backup threshold} value.
On the contrary, CBP and TBP exhibited an irregular behavior concerning different parameter values. This situation occurs since \emph{backup period} and \emph{backup task count} are fixed, and they do not completely represent the dynamics of the voltage trace.  The conclusion is that DBP is more responsive while the other two cannot adapt to the voltage trace dynamics.

Figure~\ref{fig:results_ea_counters} presents the approximated energy consumption of the simulation architecture (including backup policy blocks) concerning different backup policies. Considering the parameter values that maximize the outcome of each policy from the previous simulation (\signal{counter1\_val}), the energy consumption of the DBP is the highest among the others. This is since the counters were on and incremented more with the DBP. One can calculate the amount of energy consumption per one counter increment operation as:
\begin{align}
    E_{DBP}=1981/223=8.88 Joules/Increment, \nonumber \\ 
    E_{CBP}=1768/191=9.26 Joules/Increment, \nonumber \\ 
    E_{TBP}=1626/184=8.84 Joules/Increment. \nonumber
\end{align}
Therefore, from the energy consumption point of view, TBP is as good as the DBP for the particular voltage trace used in the simulations. Another observation is that the energy consumption trend presented in Figure~\ref{fig:results_ea_counters} for CBP and TBP is the opposite of that of DBP. The reason is that increasing the \emph{backup threshold} in DBP reduces the available computation time, and in turn, decreases the available time for the counter increment operation. However, in CTB and TBP policies, increasing the \emph{backup period} and \emph{task count} parameters delays the backup, hence the volatile counters are incremented more before a backup operation.
    
Figure~\ref{fig:results_ea_norm} presents the energy approximation of the \sysname framework. The graphs in this figure are roughly following the opposite trend of the graphs depicted in Figure~\ref{fig:results_ea_counters}. This happens because when counters are active \sysname is not, and vice versa. By considering the points where the tuning parameter maximize the COUNTER 1 value (pointed values in Fig. \ref{fig:results_ea_norm}, one for each policy), it is possible to state that the lowest energy consumption, in term of NORM usage, is achieved by DBP. This checks with the theory since the principle of DBP is to use less frequently \nvr and backup only when necessary.

As a summary of our simulations, we conclude that \sysname can be used to validate and analyze logic systems, including non-volatile memory elements. 

%% file: 6-conclusions.tex
\section{Conclusions and Future Work}
\label{sec:conclusion}
In this paper, we introduced \sysname which is an FPGA-based emulation framework. \sysname can emulate any intermittent computing system that exploits fast non-volatile memories to store temporary status in case of supply failures.  \sysname comprises auxiliary blocks that simulate the behavior of an unregular power supply, which is typical to any batteryless transient computing system powered by energy harvesters. \sysname takes into account the delay of the NVMs and approximates the energy consumption of the emulated technology. Our evaluation showed that \sysname can be used to emulate and validate a  FeRAM-based custom non-volatile digital logic successfully. We conclude that \sysname is appropriate for verifying the behavior of such new types of systems over long time scales, typical of duty-cycling energy-neutral Internet of Things (IoT) applications. 

Future studies can target the emulation of a more sophisticated non-volatile logic, such as a non-volatile processor architecture.  RISC-V processor family is freely available and a good candidate for IoT computing applications. A non-volatile RISC-V can be implemented, and its behavioral verification can be done using the proposed FPGA architecture. Moreover, the accuracy of the energy approximation block can be observed by comparing the actual ASIC implementation of the processor concerning its implementation in the proposed emulation architecture.